\documentclass[]{article}
\usepackage{latexsym}
\usepackage{amsmath}
\usepackage{amsfonts}
\usepackage{amssymb}
\usepackage{amscd}
\input xy
\font\blackital=cmmib10  \skewchar\blackital='177
\font\sblackital=cmmib7  \skewchar\sblackital='177
\font\ssblackital=cmmib5  \skewchar\ssblackital='177
\font\sanss=cmss10 \font\ssanss=cmss8 scaled 900
\font\sssanss=cmss8 scaled 600 
 \font\teneurm=eurm10 
\font\blackboard=msbm10 \font\sblackboard=msbm7
\font\ssblackboard=msbm5 \font\caligr=eusm10 \font\scaligr=eusm7
\font\sscaligr=eusm5  \font\fraktur=eufm10
\font\sfraktur=eufm7 \font\ssfraktur=eufm5 
\newfam\bbfam
\textfont\bbfam=\blackboard \scriptfont\bbfam=\sblackboard
\scriptscriptfont\bbfam=\ssblackboard

\newfam\ssfam
\textfont\ssfam=\sanss \scriptfont\ssfam=\ssanss
\scriptscriptfont\ssfam=\sssanss
\def\ss#1{{\fam\ssfam\relax#1}}

\newfam\clfam
\textfont\clfam=\caligr \scriptfont\clfam=\scaligr
\scriptscriptfont\clfam=\sscaligr
\def\cl#1{{\fam\clfam\relax#1}}

\newfam\frfam
\textfont\frfam=\fraktur \scriptfont\frfam=\sfraktur
\scriptscriptfont\frfam=\ssfraktur
\def\fr#1{{\fam\frfam\relax#1}}
\newfam\gpfam
\def\eu#1{\hbox{$\fam\gpfam\relax#1\textfont\gpfam=\teneurm$}}

\font\bsymb=cmsy10 scaled\magstep2
\def\all#1{\setbox0=\hbox{\lower1.5pt\hbox{\bsymb \char"38}}\setbox1=\hbox{$_{#1}$} \box0\lower2pt\box1}
\def\exi#1{\setbox0=\hbox{\lower1.5pt\hbox{\bsymb \char"39}}\setbox1=\hbox{$_{#1}$} \box0\lower2pt\box1}

\catcode`\"=12
\mathchardef\za="710B  
\mathchardef\zb="710C  
\mathchardef\zg="710D  
\mathchardef\zd="710E  
\mathchardef\zve="710F 
\mathchardef\zz="7110  
\mathchardef\zh="7111  
\mathchardef\zvy="7112 
\mathchardef\zi="7113  
\mathchardef\zk="7114  
\mathchardef\zl="7115  
\mathchardef\zm="7116  
\mathchardef\zn="7117  
\mathchardef\zx="7118  
\mathchardef\zp="7119  
\mathchardef\zr="711A  
\mathchardef\zs="711B  
\mathchardef\zt="711C  
\mathchardef\zu="711D  
\mathchardef\zvf="711E 
\mathchardef\zq="711F  
\mathchardef\zc="7120  
\mathchardef\zw="7121  
\mathchardef\ze="7122  
\mathchardef\zy="7123  
\mathchardef\zvp="7124  
\mathchardef\zvr="7125 
\mathchardef\zvs="7126 
\mathchardef\zf="7127  
\mathchardef\zG="7000  
\mathchardef\zD="7001  
\mathchardef\zY="7002  
\mathchardef\zL="7003  
\mathchardef\zX="7004  
\mathchardef\zP="7005  
\mathchardef\zS="7006  
\mathchardef\zU="7007  
\mathchardef\zF="7008  
\mathchardef\zC="7009  
\mathchardef\zW="700A  
\catcode`\"=\active

\def\eza{\eu\za}

\def\eze{\eu\ze}

\def\ezk{\eu\zk}
\def\ezt{\eu\zt}

\def\ezp{\eu\zp}
\def\sT{{\ss T}}
\def\cC{{\cl C}}
\def\cR{{\cl R}}

\def\g{{\fr g}}

\newtheorem{theo}{Theorem}
\newtheorem{lemma}{Lemma}
\newtheorem{coll}{Collorary}

\def\nn{\nonumber}
\def\*{{\textstyle *}}
\def\s*{{\scriptstyle *}}
\def\s{{\scriptstyle *}}
 
\def\sss#1{\hbox{\ssanss #1}}
\def\R{{\mathbb R}}

\def\ssT{\sss T}

\def\xd{\operatorname{d}\!}

\def\ix{\operatorname{i}}
\def\vta{\operatorname{v}_\zt}

\def\dt{\xd_{\sss T}}
\def\vt{\textsf{v}_{\sss T}}

\def\ix{\operatorname{i}}
\def\ot{\otimes}
\def\ti{\times}
\def\pa{\partial}
\def\bv{\bar{v}}
\def\dx{\dot{x}}
\def\ra{\rightarrow}

\def\wt{\widetilde}

\def\sP{{\ss P}}

\newcommand{\be}{\begin{equation}}
\newcommand{\ee}{\end{equation}}

\newcommand{\bea}{\begin{eqnarray}}
\newcommand{\eea}{\end{eqnarray}}
\newcommand{\beas}{\begin{eqnarray*}}
\newcommand{\eeas}{\end{eqnarray*}}

\xyoption{all} \tolerance=500 \textwidth15.6cm \textheight23cm
\begin{document}
\title{Geometrical Mechanics on algebroids\thanks{Research
supported by the Polish Ministry of Scientific Research and
Information Technology under the grant No. 2 P03A 036 25.}}

        \author{
        Katarzyna  Grabowska$^1$, Janusz Grabowski$^2$, Pawe\l\ Urba\'nski$^1$\\
        \\
         $^1$ {\it Division of Mathematical Methods in Physics}\\
                {\it University of Warsaw} \\
         $^2$ {\it Institute of Mathematics}\\
                {\it Polish Academy of Sciences}
                }
\date{}
\maketitle
\begin{abstract}
A natural geometric framework is proposed, based on ideas of
   W.~M.~Tulczyjew, for constructions of dynamics on general algebroids. One obtains formalisms
   similar to the Lagrangian and the Hamiltonian ones. In contrast with recently studied
concepts of Analytical Mechanics on Lie algebroids, this approach
requires much less than the presence of a Lie algebroid structure
on a vector bundle, but it still reproduces the main features of
the Analytical Mechanics, like the Euler-Lagrange-type equations,
the correspondence between the Lagrangian and Hamiltonian
functions (Legendre transform) in the hyperregular cases, and a
version of the Noether Theorem.

\bigskip\noindent
\textit{MSC 2000: 70G45, 70H03, 53C99, 53D17.}

\medskip\noindent
\textit{Key words: Lie algebroid, double vector bundle, Lagrangian function,
Euler-Lagrange equations.}
\end{abstract}

\section{Introduction}
Lie algebroids have been introduced repeatedly into differential
geometry since the early 1950s, and also into physics and algebra,
under a wide variety of names. They have been also recognized as
infinitesimal objects for Lie groupoids by J.~Pradines \cite{Pr}.
We refer to \cite{Ma} for basic definitions, examples, and an
extensive list of publications in this area. The classical
Cartan differential calculus on a manifold $M$, including the
exterior derivative $\xd\,$, the Lie derivative $\pounds$, etc.,
can be viewed as being associated with the canonical Lie algebroid
structure on $\sT M$ represented by the Lie bracket of vector
fields, and therefore it has obvious generalizations to an
arbitrary Lie algebroid.

The tangent bundle $\sT M$ is of course a canonical Lie algebroid.
It is associated with the canonical Poisson tensor (symplectic
form) on $\sT^\*M$. Other canonical objects associated with $\sT
M$ are: the canonical isomorphism
    $$ \eza_M\colon \sT\sT^\* M \longrightarrow  \sT^\* \sT M
                              $$
of double vector bundles, discovered by Tulczyjew \cite{Tu}, that is the dual
to the well-known flip
    $$ \ezk_M \colon \sT\sT M\longrightarrow \sT \sT M,
                              $$ and the tangent lift $\dt$ of tensor
fields on $M$ to tensor fields on $\sT M$ (cf. \cite{YI, PT,
GU1}).

        Being related to many areas of geometry, like connection theory,
cohomology theory, invariants of foliations and pseudogroups,
symplectic and Poisson geometry, etc., Lie algebroids became
recently an object of extensive studies. In \cite{We},
A.~Weinstein had posed the problem of finding a framework for
Analytical Mechanics based on a general Lie algebroid. This task
was undertaken in \cite{LMM, Li, Mar1, Mar2} to get a Lie
algebroid version of the geometric construction of the
Euler-Lagrange equations due to J. Klein \cite{Kl}. In this
approach the E-L equation is a first order equation on the Lie
algebroid bundle $E$ with some compatibility conditions. In the
classical version ($E= \sT M$), the Klein's method is based on the
vector bundle structure of $\sT M$ and the existence of a
vector-valued 1-form, so called the 'soldering form'. Such a form
does not exist for a general Lie algebroid and the conclusion is
that the immediate analogy for the Klein's approach does not exist
as well. In a series of papers E. Mart\'\i nez has  proposed an
interesting modified version of the Klein's method, in which the
bundles tangent to $E$ and $E^\*$ are replaced by the
prolongations (in the sense of Higgins and Mackenzie \cite{HM}) of
$E$ with respect to the vector bundle projections $\zt\colon E
\rightarrow M$ and $\zt^\*\colon E^\* \rightarrow M$. A similar
approach for structures more general than Lie algebroids has been
proposed by M.~Popescu and P.~Popescu \cite{PP}.

The ideas of J.~Klein go back to 1962. Since then a lot of work
has been done to get a better understanding of the geometric
background for Analytical Mechanics. In the papers of Tulczyjew
\cite{Tu2} and de Le\'on with Lacomba \cite{LL} we find another
geometric constructions of the E-L equation. The starting point
for the Tulczyjew's construction is the dynamics of a system, i.e.
a Lagrangian submanifold of $\sT\sT^\* M$, which is the inverse
image of $\xd L (M)$ with respect to the canonical diffeomorphism
$\za_M\colon \sT\sT^\* M \rightarrow \sT^\*\sT M$, where $L$ is a
function (Lagrangian) on $M$. The diffeomorphism $\za_M$, or its
dual $\zk_M \colon \sT\sT M \rightarrow  \sT \sT M$, represent,
unlike the 'soldering form', the complete structure of the tangent
bundle.

On the other hand, in a couple of papers \cite{GU3, GU2}, two of
us have developed an approach to Lie algebroids based on the
analogue of these canonical diffeomorphisms, which form a part  of
the so called {\it Tulczyjew triple}, and which appear to be
morphisms of double vector bundles. This allowed us to introduce
the notion of a (general, not necessarily Lie) {\it algebroid} as
a morphism of certain double vector bundles or, equivalently, as a
vector bundle equipped with a linear 2-contravariant tensor.

        What we propose in this paper is to adopt the Tulczyjew  approach
        \cite{Tu} (cf. also \cite{TU}) to the case of a general
algebroid. In particular, we obtain a geometric construction of an
equation which was suggested by A.~Weinstein as a Lie algebroid
version of the Euler-Lagrange equation. The main difference with
the papers like \cite{LMM, Mar1, Mar2, PP} is not only that we
deal with general algebroids but also that in this case we avoid
objects like 'soldering form', symplectic algebroids,
Poincar\'e-Cartan sections, Lie algebroid prolongations, etc.

    We postpone to a separate publication the  discussion of the calculus of
    variations for a general Lie algebroid. It requires more fundamental discussion
    on the virtual displacements and it is closely related to the problem of
    integrability of algebraic structures in the sense in which Lie
    algebroids can be integrated to Lie groupoids.

The paper is organized as follows.
First we fix the notation. In the next two sections, the classical
Tulczyjew triple and rudiments of the approach from \cite{GU3} and
\cite{GU2} on viewing Lie algebroids as double bundle morphisms
are presented. In Section 3 we recall the definition of the
complete lift of tensor fields. In Section 4 we present the
concepts of deriving dynamics on general algebroids $E$, similar
to that in the Lagrangian and Hamiltonian formalisms. The
equations we discuss are equations for paths in $E^\*$ and $E$.
The latter are often regarded as algebroid generalizations of the
Euler-Lagrange equations. A variant of the Noether Theorem is
proved.
Section 5 contains a couple of examples which generalize the geodesic
and the Wong equations.

\subsection{Notation}
        Let $M$ be a smooth manifold. We denote by $\ezt_M \colon \sT M
\rightarrow M$ the tangent vector bundle and by $\ezp_M \colon
\sT^\* M\rightarrow M$ the cotangent vector bundle.

        Let $\zt\colon E \rightarrow M$ be a vector bundle and let $\zp
\colon E^\* \rightarrow M$ be the dual bundle. We use the
following notation for tensor bundles:
                $$
        \zt^{\otimes k}\colon E\otimes_ M \cdots \otimes _M E  = \otimes
^k_M(E) \longrightarrow M
             $$
       and the module of sections over $\cC^\infty(M)$:
                $$
        \otimes ^k(\zt)= \zG(\otimes ^k_M(E)).
                 $$
  By $\langle \cdot ,\cdot \rangle $, we denote the canonical pairing
between $E$ and $E^\*$ as well as pairings between  the
corresponding  tensor bundles, e.g.,
    $$ \langle \cdot ,\cdot \rangle \colon \otimes ^k_M (E) \times_M
\otimes ^k_M (E^\*) \longrightarrow \R,
                              $$
  and pairings of sections, e.~g.,
    $$ \langle \cdot ,\cdot \rangle \colon \otimes ^k (\zt) \times
\otimes ^k (\zp) \longrightarrow \cC^\infty (M).
                              $$
  Let  $K$ be a section of the tensor bundle $\otimes ^k_M(E)$, $K \in
\otimes ^k(\zt)$. We denote by $\zi (K)$ the corresponding linear
function on the dual bundle
    \beas
\zi(K) &\colon& \otimes^ k_M(E^\*) \rightarrow  \R \\
    &\colon& a  \mapsto \langle K(m), a\rangle, \ \ \  m =
  \zp^{\otimes k}(a).\eeas
  For a section $X$ of $\zt$ ($X\in \otimes ^1(\zt) $), we have the usual
operator of insertion
    \beas
\ix_X  &\colon& \otimes ^{k+1}(\zp)\rightarrow \otimes ^k(\zp)  \\
  &\colon& \zm_1\otimes \cdots \otimes \zm_{k+1} \mapsto \langle X,
\zm_1\rangle  \zm_2\otimes \cdots \otimes \zm_{k+1}.
                    \eeas
Let $\zL\in \otimes ^2(\zt)$. We denote by $\widetilde{\zL}$ the
mapping
    $$ \widetilde{\zL}\colon E^\* \rightarrow E, \ \
\widetilde{\zL}\circ \zm = \ix_\zm \zL.
                              $$
By $\zD_E$ we denote the Liouville (called also Euler) vector
field on the vector bundle $E$.

  \subsection{Local coordinates}
  Let $(x^a), \ a=1,\dots,n$, be a coordinate system in $M$. We introduce
the induced coordinate systems
    \beas
    (x^a, {\dot x}^b) &\ \text{in} \ \sT M,\\
    (x^a, p_b) &\ \text{in} \ \sT^\* M.
                    \eeas
  Let $(e_1,\dots,e_m)$  be a basis of local sections of $\zt\colon
E\rightarrow M$ and let $(e^{1}_*,\dots, e^{m}_*)$ be the dual
basis of local sections of $\zp\colon E^\*\rightarrow M$. We have
the induced coordinate systems:
    \beas
    (x^a, y^i),\quad & y^i=\zi(e^{i}_*), \quad \text{in} \ E,\\
    (x^a, \zx_i), \quad &\zx_i = \zi(e_i),\quad \text{in} \ E^\* ,
    \eeas
  and 
    \beas
    (x^a, y^i,{\dot x}^b, {\dot y}^j ) &  \quad \text{in} \ \sT E,\\
    (x^a, \zx_i, {\dot x}^b, {\dot \zx}_j) & \quad \text{in} \ \sT E^\* ,\\
    (x^a, y^i, p_b, \zp_j) & \quad \text{in}\ \sT^\*E, \\
    (x^a, \zx_i, p_b, \zf^j) & \quad \text{in}\ \sT^\* E^\* .
    \eeas
The adapted coordinates on the above bundles define canonical {\it
double vector bundle} structures on them in the sense of
J.~Pradines (\cite{Pr1,Pr2}, cf. also \cite{BM,Ma1}).
    The Liouville vector field has the form
  $\zD_E=y^i\partial_{y^i}$.
We have the canonical symplectic forms:
    $$ \zw_{E^*} = \xd p_a\wedge \xd x^a +\xd \zf^i \wedge \xd \zx_i
                              $$
on $\sT^\*E^\*$ and
    $$ \zw_{E} = \xd p_a\wedge \xd x^a +\xd \zp_i \wedge \xd y^i
                              $$
  on $\sT^\*E$, and the corresponding Poisson tensors
    $$ \zL_{E^*} = \partial _{p_a}\wedge \partial _{x^a} + \partial
_{\zf^j} \wedge \partial _{\zx_j}\ \ \text{and}\ \
  \zL_{E} = \partial _{p_a}\wedge \partial _{x^a} + \partial
_{\zp_j} \wedge \partial _{y^j}.
                              $$
There is also a canonical isomorphism (cf. \cite{Du, KU, GU2})
    $$\cR_\zt \colon \sT^\*E^\* \longrightarrow \sT^\* E
                    $$
  being an anti-symplectomorphism and also an isomorphism of double
vector bundles:
$$\xymatrix{
 & \sT^\ast E^\ast \ar[rrr]^{\cR_r} \ar[dr]^{\sT^\ast\pi}
 \ar[ddl]_{\pi_{E^\ast}}
 & & & \sT^\ast E\ar[dr]^{\pi_E}\ar[ddl]_/-20pt/{\sT^\ast\tau}
 & \\
 & & E\ar[rrr]^/-20pt/{id}\ar[ddl]_/-20pt/{\tau}
 & & & E \ar[ddl]_{\tau}\\
 E^\ast\ar[rrr]^/-20pt/{id}\ar[dr]^{\pi}
 & & & E^\ast\ar[dr]^{\pi} & &  \\
 & M\ar[rrr]^{id}& & & M &
}$$ In local coordinates, $\cR_\zt$ is given by
    $$ (x^a,y^i, p_b, \zp_j)\circ \cR_\zt = (x^a, \zf^i, -p_b,\zx_j).
                              $$
\subsection{Classical Tulczyjew triple}

    Let $M$ be the {\it configuration manifold} of a mechanical system.  The
cotangent bundle $\sT^\*M$ is the {\it phase space} of the system.
Elements of the phase space are {\it momenta}.  The commutative
diagram

$$\xymatrix@C-10pt{
\sT^\ast\sT^\ast M \ar[dr]_{\pi_{\sT^\ast M}}
 &  &
\sT \sT^\ast M \ar[ll]_{\beta_{(\sT^\ast M,\omega_M)}}
\ar[rr]^{\alpha_M} \ar[dl]_{\tau_{\sT^\ast M}} \ar[dr]^{\sT\pi_M}
 &  &
\sT^\ast\sT M \ar[dl]^{\pi_{\sT M}} \\
\quad  & \sT^\ast M \ar[dr]^{\pi_M} & \quad& \sT M \ar[dl]_{\tau_M} &\quad  \\
 & & M & &}
$$

\noindent known as the Tulczyjew triple, contains the geometric structures
used to formulate the dynamics of
the system.  The dynamics is a differential equation $D \subset
\sT\sT^\*M$.  A solution $\zg \colon I \rightarrow \sT^\*M$ of
this equation is a phase space trajectory of the system.
Trajectories of the system in the configuration manifold $M$ are solutions of the
second-order Euler-Lagrange equation
        $$E_L = \sT^2\zp_M(\sP D),$$
where the set
        $$\sP D =\sT D\cap\sT^2\sT^\* M\subset \sT^2\sT^\* M$$
is a second order differential equation called the {\it prolongation} of $D$.
Note however that in general these trajectories in $M$ do not
determine $D$.

We have recognized the presence of a canonical symplectic structure in
$\sT\sT^\*M$ with the symplectic form $\xd_T\zw_M$.  In most cases
of interest in physics the dynamics is a Lagrangian
submanifold of $(\sT\sT^\*M,\xd_T\zw_M)$.  Morphisms $\za_M$ and
$\zb_{(\sT^\*M,\zw_M)}$ are canonical symplectomorphisms from
$(\sT\sT^\*M,\xd_T\zw_M)$ to $(\sT^\*\sT M,\zw_{\sT M})$ and to
$(\sT^\*\sT^\*M,\zw_{\sT^\*M})$.  These symplectomorphisms with
cotangent bundles create the possibility of generating the
dynamics from (generalized) Lagrangians associated with $\sT M$ or
(generalized) Hamiltonians associated with $\sT^\*M$ (cf.
\cite{Tu, TU, Ur}).

\section{Algebroids as double vector bundle morphisms\label{S1}}
It is well known that Lie algebroid structures on the vector
bundle $E$ correspond to linear Poisson structures on $E^\*$. A
2-contravariant tensor $\zL$ on $E^\*$ is called {\it linear} if
the corresponding mapping $\widetilde{\zL} \colon \sT^\* E^\*
\rightarrow \sT E^\*$ is a morphism of double vector bundles. This
is the same as to say that the corresponding bracket of functions
is closed on (fiber-wise) linear functions. The commutative
diagram
$$\xymatrix{
\sT^\ast E^\ast\ar[r]^{\widetilde\Lambda} \ar[d]_{\cR_\tau} & \sT E^\ast \\
\sT^\ast E\ar[ur]^{\ze} & }
$$
describes a one-to-one correspondence between linear
2-contravariant tensors $\zL$ on $E^\*$ and homomorphisms of
double vector bundles covering the identity on $E^\*$ (cf.
\cite{KU, GU2}).
\be\xymatrix{
 & \sT^\ast E \ar[rrr]^{\varepsilon} \ar[dr]^{\pi_E}
 \ar[ddl]_{\sT^\ast\tau}
 & & & \sT E^\ast\ar[dr]^{\sT\pi}\ar[ddl]_/-20pt/{\tau_{E^\ast}}
 & \\
 & & E\ar[rrr]^/-20pt/{\varepsilon_r}\ar[ddl]_/-20pt/{\tau}
 & & & \sT M \ar[ddl]_{\tau_M}\\
 E^\ast\ar[rrr]^/-20pt/{id}\ar[dr]^{\pi}
 & & & E^\ast\ar[dr]^{\pi} & &  \\
 & M\ar[rrr]^{id}& & & M &
}\label{F1.3}\ee The core of a double vector bundle is the
intersection of the kernels of the projections. It is obvious that
the core of $\sT^\* E$ (resp., $\sT E^\*$) can be identified with
$\sT^\* M$ (resp., $E^\*$). With these identifications the induced
by $\ze$ morphism of cores is a morphism
$$\ze_c \colon \sT^\* M \rightarrow E^\*.$$
In local coordinates, every  $\ze$ as in (\ref{F1.3}) is of the form
                \be (x^a, \zx_i, {\dot x}^b, {\dot \zx}_j) \circ \ze = (x^a,
\zp_i, \zr^b_k(x)y^k, c^k_{ij}(x) y^i\zp_k + \zs^a_j(x) p_a)
                                                             \label{F1.4}\ee
and it corresponds to the linear tensor on $E^\*$
                \be \zL_\ze = c^k_{ij}(x)\zx_k \partial _{\zx_i}\otimes \partial
_{\zx_j} + \zr^b_i(x) \partial _{\zx_i} \otimes \partial _{x^b} -
\zs^a_j(x) \partial _{x^a} \otimes \partial _{\zx_j}.
                                                             \label{F1.5}\ee
We have also
\beas
                (x^a, \dot x^b)\circ \ze_r &= (x^a, \zr^b_k(x)y^k), \\
                (x^a,  \zx_i)\circ \ze_c &= (x^a, \zs^b_i(x)p_b).
                                                                \eeas
In \cite{GU2} by  {\it algebroids} we meant the morphisms
(\ref{F1.3}) of double vector bundles covering the identity on
$E^\*$, while {\it Lie algebroids} were those algebroids for which
the tensor $\zL_\ze$ is a Poisson tensor. The relation to the
canonical definition of Lie algebroid is given by the following
theorem.

\begin{theo}{\em \cite{GU3, GU2}}        \label{C1.3}
An algebroid structure $(E,\ze)$ can be equivalently defined as a
bilinear bracket $[\cdot ,\cdot]_\ze $ on sections of $\zt\colon
E\rightarrow M$, together with vector bundle morphisms $a^\ze_l,\,
a^\ze_r \colon E\rightarrow \sT M$ (left and right anchors), such
that
                $$ [fX,gY]_\ze = f(a^\ze_l\circ X)(g)Y -g(a^\ze_r \circ Y)(f) X
+fg [X,Y]_\ze
                                                   $$
         for $f,g \in \cC^\infty (M)$, $X,Y\in \otimes ^1(\zt)$.
The bracket and anchors are related to the 2-contravariant tensor $\zL_\ze$ by
the formulae
\beas
        \zi([X,Y]_\ze)&= \{\zi(X), \zi(Y)\}_{\zL_\ze},  \\
        \zp^\*(a^\ze_l\circ X(f))       &= \{\zi(X), \zp^\*f\}_{\zL_\ze}, \\
        \zp^\*(a^\ze_r\circ X(f))       &= \{\zp^\* f, \zi(X)\}_{\zL_\ze}.
                                                   \eeas
        We have also $a^\ze_l=\ze_r$ and $a^\ze_r = (\ze_c)^\*$.
        The algebroid $(E,\ze)$ is a Lie algebroid if and only if the tensor
$\zL_\ze$ is a Poisson tensor.
\end{theo}
To every algebroid $\ze$ there is the {\it adjoint algebroid}
$\ze^+:\sT^\s E\ra\sT E^\s$ which is the bundle morphism dual to
$\ze$ with respect to the  projections onto $E^\s$ and which
corresponds to the transposition of the tensor $\zL_\ze$. The
algebroid is {\it skew-symmetric} if and only if $\ze^+=-\ze$. An
algebroid we call a (left or right) {\it connection} if one of the
anchors (right or left, respectively) is trivial. In this case we
have for the bracket $\nabla^\ze_XY=[X,Y]_\ze$ associated with,
say, a left connection $\ze$, the standard properties of a linear
connection: $\nabla_{fX}^\ze Y=f\nabla_{X}^\ze Y$ and
$\nabla_{X}^\ze fY=f\nabla_{X}^\ze Y+(a^\ze_l\circ X)(f)Y$.

The canonical example of a mapping $\ze$ in the case of $E=\sT M$ is
given by $\ze = \eze_M = \eza^{-1}_M$ -- the inverse to the
Tulczyjew isomorphism $\eza_M$ that can be defined as the dual to the
isomorphism of double vector bundles
$$
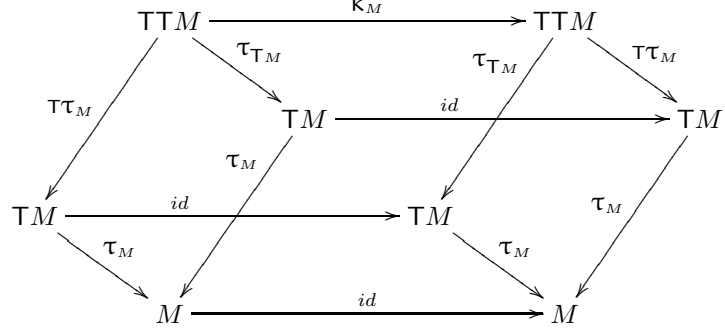
\xymatrix{
 & \sT \sT M \ar[rrr]^{\ezk_M} \ar[dr]^{\ezt_{\ssT M}}
 \ar[ddl]_{\sT\ezt_M}
 & & & \sT \sT M\ar[dr]^{\sT\ezt_M}\ar[ddl]_/-20pt/{\ezt_{\ssT M}}
 & \\
 & & \sT M\ar[rrr]^/-20pt/{id}\ar[ddl]_/-20pt/{\ezt_M}
 & & & \sT M \ar[ddl]_{\ezt_M}\\
 \sT M\ar[rrr]^/-20pt/{id}\ar[dr]^{\ezt_M}
 & & & \sT M\ar[dr]^{\ezt_M} & &  \\
 & M\ar[rrr]^{id}& & & M &
}$$
In general, the algebroid structure map $\ze$ is not an isomorphism and,
consequently, its dual $\zk^{-1} = \ze^{\s_r}$  with respect to
the right projection is a relation and not a mapping.

\section{The algebroid lift $\dt^\ze$} For a tensor field $K\in
\otimes ^k(\zt)$, we can define the vertical lift
$\textsf{v}_\zt(K) \in \otimes ^k(\ezt_E)$ (cf. \cite{GU1,YI}). In
local coordinates,
\be \textsf{v}_\zt(f^{i_1\dotsi_k} e_{i_1}\otimes \cdots
\otimes e_{i_k}) = f^{i_1\dotsi_k} \partial _{y^{i_1}} \otimes
\cdots \otimes \partial _{y^{i_k}}. \label{F1.14}\ee
A particular case of the vertical lift is the lift $\vt(K)$ of a
contravariant tensor field $K$ on $M$ into a contravariant tensor
field on $\sT M$.
It is well  known (see \cite{YI,GU1}) that in the case of
$E=\sT M$ we have also the tangent lift $\dt \colon \otimes
(\ezt_M) \rightarrow \otimes (\ezt_{\ssT M})$ which is a
$\vt$-derivation. It turns out that the presence of such a lift
for a vector bundle is equivalent to the presence of an algebroid
structure. Note first that we can extend $\ze$ naturally to
mappings (cf. \cite{GU3, GU2})
$$ \ze^{\otimes r} \colon \otimes ^r_E \sT^\* E \longrightarrow
\sT \otimes ^r_M E^\*,\quad r\ge 0.
$$

\begin{theo}{\em \cite{GU3, GU2}}            \label{C1.8}
Let $(E,\ze)$ be an algebroid. For $K\in \otimes ^k(\zt)$, $k\ge
0$, the equality
\be \zi(\dt^\ze(K)) = \dt (\zi(K))\circ \ze^{\otimes k} \label{F1.17}\ee
defines the  tensor field $\dt^\ze(K) \in \otimes ^k(\ezt_E)$  which
is linear and the mapping
$$ \dt^\ze \colon \otimes (\zt) \longrightarrow \otimes (\ezt_E) $$
is a $\vta$-derivation of degree $0$. In local coordinates,
\be
\begin{array}{l}\vspace{5pt}\dt^\ze(f(x)) =  y^i \zr^a_i(x) \frac{\partial f}{\partial
x^a}(x),\\
\dt^\ze(f^i(x)e_i) = f^i(x)\zs^a_i(x)\partial _{x^a} +
\left( y^i \zr^a_i(x) \frac{\partial f^k}{\partial x^a}(x) +
c^k_{ij}(x)y^if^j(x) \right) \partial _{y^k}.\end{array}\label{F1.18}\ee
Conversely, if $D\colon  \otimes (\zt)\rightarrow \otimes (\ezt_E)$ is
a $\vta$-derivation of degree $0$ such that $D(K)$ is linear for
each $K \in \otimes ^1(\zt)$, then there is an algebroid structure
$\ze$ on $\zt\colon E\rightarrow M$ such that $D = \dt^\ze$.
\end{theo}

\begin{theo}{\em \cite{GU2}}                          \label{C1.14}
Let $\ze$ be an algebroid structure on $\zt\colon E\rightarrow M$. The
following properties of $\ze$ are equivalent:
\begin{description}
        \item[(a)] $\ze$ is a Lie algebroid structure,
        \item[(b)] $\zL_\ze$ is a Poisson tensor,
        \item[(c)] $\zL_E$ and $\dt\zL_\ze$ are $\ze$--related,
        \item[(d)] $\dt^\ze([X,Y]_\ze) = [\dt^\ze(X), \dt^\ze(Y)]$ for all
$X,Y\in \otimes ^1(\zt)$.
\end{description}
\end{theo}

\section{Lagrangian and Hamiltonian formalisms for general
algebroids} The double vector bundle morphism (\ref{F1.3}) can be
extended to the following algebroid analogue of the Tulczyjew
triple
\be\xymatrix@C-5pt{
 &\sT^\ast E^\ast \ar[rrr]^{\widetilde{\Lambda}}
\ar[ddl]_{\pi_{E^\ast}} \ar[dr]^{\sT^\ast\pi}
 &  &  & \sT E^\ast \ar[ddl]_/-25pt/{\ezt_{E^\ast}} \ar[dr]^{\sT\pi}
 &  &  & \sT^\ast E \ar[ddl]_/-25pt/{\sT^\ast\ezt} \ar[dr]^{\pi_E}
\ar[lll]_{\varepsilon}
 & \\
 & & E \ar[rrr]^/-20pt/{\varepsilon_r}\ar[ddl]_/-20pt/{\ezt}
 & & & \sT M\ar[ddl]_/-20pt/{\ezt_M}
 & & & E\ar[lll]_/+20pt/{\varepsilon_r}\ar[ddl]_{\ezt}
 \\
 E^\ast\ar[rrr]^/-20pt/{id} \ar[dr]^{\pi}
 & & & E^\ast\ar[dr]^{\pi}
 & & & E^\ast\ar[dr]^{\pi}\ar[lll]_/-20pt/{id}
 & & \\
 & M\ar[rrr]^{id}
 & & & M & & & M\ar[lll]_{id} &
}\label{F1.3b}\ee
The left-hand side is Hamiltonian, the right-hand side is
Lagrangian, and the 'dynamics' lives in the middle.

Any Lagrangian function $L:E\ra\R$ defines a Lagrangian
submanifold $N=(\xd L)(E)$ in $\sT^\* E$, being the image of the
de Rham differential $\xd L$, i.e. the image of the section $\xd
L:E\ra\sT^\* E$. The further image $D=\ze(N)$ can be understood as
an implicit differential equation on $E^\*$, solutions of which
are `phase trajectories' of the system. The Lagrangian defines
also smooth maps: $L_{eg}:E\ra E^\*$ and $\wt{L}_{eg}:E\ra \sT
E^\*$ by
$$L_{eg}=\zt_{E^\*}\circ\ze\circ\xd L=\sT^\*\zt\circ\xd L$$ and
$\wt{L}_{eg}=\ze\circ\xd L$. The map $L_{eg}$  is {\it de facto}
the vertical derivative of $L$ and  is the analogue of the {\it
Legendre mapping}, incorrectly interpreted  by many authors as the
Legendre transformation associated with $L$ (as the Legendre
transformation is  the passage from a Lagrangian to a Hamiltonian
generating object as explained in \cite{TU}). The introduced
ingredients produce implicit differential equations, this time for
curves $\zg:I\ra E$.

The first equation, which will be
denoted by $(E^1_L)$, is represented by the inverse image
\be\label{first}E_L^1=\sT(L_{eg})^{-1}(D)\ee
of $D$ with respect to the derivative
$\sT(L_{eg}):\sT E\ra\sT E^\*$ of $L_{eg}:E\ra E^\*$. This simply
means that $\zg:I\ra E$ is a solution $(E^1_L)$ if and only if
$L_{eg}\circ\zg$ is a solution of $(D)$, i.e.
$\sT(L_{eg}\circ\zg)\subset D$. This construction corresponds to
that of de L\'eon and Lacomba \cite{LL}.

The second equation, which will be denoted by $(E^2_L)$, is defined by
\be E^2_L = \{v\in \sT E \colon \sT(\ze\circ \xd L)(v) \in \sT^2 E^\*
\},\label{F1.4ab}\ee
where $\sT^2 E^\* \subset \sT\sT E^\*$ is the subset of holonomic vectors, i.e. such that
$\zt_{\sT E^\*}(w) = \sT\zt_{E^\*} (w)$.
It follows from the commutativity of the diagram (\ref{F1.3b}) that
$$ E^2_L = \{v\in \sT E \colon \ze\circ\zt_{\sT^\* E}(\sT\xd L(v)) =
    \sT(\sT^\*\zt)( \sT\xd L(v)) \}.$$
It is clear that this definition can be extended to any subset of $\sT^* E$.
The solutions of the second equation,  are such paths $\zg:I\ra E$
that the tangent prolongation $\sT(L_{eg}\circ\zg)$ of
$L_{eg}\circ\zg$ is exactly $\wt{L}_{eg}\circ\zg$. Since the path
$\wt{L}_{eg}\circ\zg$ belongs to $D$ by definition,
$(E^2_L)\subset(E^1_L)$. It is easy to see that the equation
$(E^2_L)$ is represented by the inverse image
$$E_L^2=\sT(\wt{L}_{eg})^{-1}(\sT^2E^\*)$$ of the subbundle
$\sT^2E^\*$ of holonomic vectors in $\sT\sT E^\*$ with respect to
the derivative $\sT(\wt{L}_{eg}):\sT E\ra\sT\sT E^\*$ of
$\wt{L}_{eg}:E\ra\sT E^\*$. In local coordinates, $D$ has the
parametrization by $(x^a,y^k)$ in the form (cf. (\ref{F1.4}))
\be\wt{L}_{eg}(x^a,y^i)= (x^a,\frac{\partial L}{\partial
y^i}(x,y), \zr^b_k(x)y^k, c^k_{ij}(x) y^i\frac{\partial
L}{\partial y^k}(x,y) + \zs^a_j(x)\frac{\partial L}{\partial
x^a}(x,y)) \label{F1.4a}\ee and the equation $(E^2_L)$, for
$\zg(t)=(x^a(t),y^i(t))$, reads \be (E^2_L):\qquad\qquad\frac{\xd
x^a}{\xd t}=\zr^a_k(x)y^k, \qquad \frac{\xd}{\xd
t}\left(\frac{\partial L}{\partial y^j}\right)= c^k_{ij}(x)
y^i\frac{\partial L}{\partial y^k}(x,y) + \zs^a_j(x)\frac{\partial
L}{\partial x^a}(x,y),\label{EL2}\ee in the full agreement with
\cite{LMM, Mar1, Mar2, We}, if only one takes into account that,
for Lie algebroids, $\zs^a_j=\zr^a_j$. As one can see from
(\ref{EL2}), the solutions are automatically admissible curves in
$E$, i.e. the velocity  $\frac{\xd }{\xd t}(\zt\circ\zg)(t)$ is
$\ze_r(\zg(t))$.

The equation $(E^1_L)$ is weaker and reads
\be(E^1_L):\qquad\qquad\frac{\xd x^a}{\xd t}=\zr^a_k(x)y_0^k, \qquad
\frac{\xd}{\xd t}\left(\frac{\partial L}{\partial
y^j}(x,y)\right)= c^k_{ij}(x) y_0^i\frac{\partial L}{\partial
y^k}(x,y_0) + \zs^a_j(x)\frac{\partial L}{\partial
x^a}(x,y_0),\label{EL1} \ee for certain choice of $(x,y_0)\in E$
(associated with $(x,y)$) satisfying $L_{eg}(x,y_0)=L_{eg}(x,y)$,
i.e. for $(x,y_0)$ such that
$$\frac{\partial L}{\partial y^i}(x,y)=\frac{\partial L}{\partial
y^i}(x,y_0).$$ Of course, when the Lagrangian is {\it
hyperregular}, i.e. when $L_{eg}:E\ra E^\*$ is a diffeomorphism,
the equations $(E^1_L)$ and $(E^2_L)$ coincide. \vskip20pt

\noindent We have also the following variant of the Noether Theorem.

\begin{theo} \label{noether} If $X$ is a section of $E$
and $f$ is a function on $M$, then
\be\pounds_{\dt^\ze(X)}L=\dt^\ze(f) \label{noether1}\ee
if and only if the function $(\zi(X)-\textsf{v}_\zp(f))\circ
L_{eg}$ on $E$ is a first integral of the equation $(E^2_L)$. In
particular, if (\ref{noether1}) is satisfied, then
$(\zi(X)-\textsf{v}_\zp(f))\circ L_{eg}$ is a first integral
of the equation $(E^1_L)$.
\end{theo}

\noindent{\bf Proof:} $(\zi(X)-\textsf{v}_\zp(f))\circ L_{eg}$ is
a first integral of $(E^2_L)$ $\Leftrightarrow$
$(\zi(X)-\textsf{v}_\zp(f))$ is a first integral of the implicit
differential equation on $E^\*$ defined by $D$ $\Leftrightarrow$
$\dt(\zi(X)- \textsf{v}_\zp(f))$ vanishes on $D$ $\Leftrightarrow$
$\dt(\zi(X)-\textsf{v}_\zp(f))\circ\ze$ vanishes on $N$. Writing
$\textsf{v}_\zp(f)$ as $\zi(f)$ we have, according to
(\ref{F1.17}),
$$\dt(\zi(X)-\textsf{v}_\zp(f))\circ\ze=\zi(\dt^\ze(X))
-\zi(\dt^\ze(f))
$$
and
$$(\zi(\dt^\ze(X))-\zi(\dt^\ze(f)))(\xd L(e))=
\pounds_{\dt^\ze(X)}L(e)-\dt^\ze(f)(e).$$
\hfill{$\blacksquare$}

\vskip10pt\noindent Note that the tensor $\zL_\ze$ gives rise also
to kind of a Hamiltonian formalism (cf. \cite{OPB}). In \cite{GU2}
and \cite{OPB} one refers to a 2-contravariant tensor as to a {\it
Leibniz structure}, that however may cause some confusion with the
{\it Leibniz algebra} in the sense of J.-L.~Loday as a
non-skew-symmetric analog of a Lie algebra. Anyhow, in the
presence of $\zL_\ze$, by the {\it hamiltonian vector field}
associated with a function $H$ on $E^\*$ we understand the
contraction $\ix_{\xd H}\zL_\ze$. Thus the question of the
Hamiltonian description of the dynamics $D$ is the question if $D$
is the image of a Hamiltonian vector field. (Of course, one can
also try to extend such a Hamiltonian formalism to more general
generating objects like Morse families.) Every such a function $H$
we call a {\it Hamiltonian  associated with the Lagrangian} $L$.
However, it should be stressed that, since $\ze$ and $\zL_\ze$ can
be degenerated, we have much more freedom in choosing generating
objects (Lagrangian and Hamiltonian) than in the symplectic case.
For instance, the Hamiltonian is defined not up to a constant but
up to a Casimir function of the tensor $\zL_\ze$ and for the
choice of the Lagrangian we have a similar freedom. However, in
the case of a hyperregular Lagrangian we recover the standard
correspondence between Lagrangians and Hamiltonians. Let us start
with the following lemma which can be easily proved exactly like
in the classical case and which reflects the fact that this
correspondence is, in principle, independent on the algebroid
structure on $E$ but which comes directly from the isomorphism
$\cR_\zt$.

\begin{lemma} \label{LH}
If the Lagrangian $L$ is hyperregular, then the Lagrange submanifold
        $N=\xd L(M)$ in $\sT^\* E$ corresponds under the canonical isomorphism
$\cR_\zt$ to the Lagrange submanifold $\xd H(M)$ in $\sT^\* E^\*$,
where $H=(\Delta_E(L)-L)\circ L_{eg}^{-1}$
\end{lemma}

\begin{coll} \label{ham}
If the Lagrangian $L$ is hyperregular, then the function
$H=(\Delta_E(L)-L)\circ L_{eg}^{-1}$ is a Hamiltonian associated
with $L$.
\end{coll}

\noindent{\bf Proof:} Since $\ze=\widetilde{\zL_\ze}\circ\cR_\zt^{-1}$, in view
of the above Lemma, $\ze(N)=\widetilde{\zL_\ze}(\xd H(M))$ which
means exactly that $\ze(N)$ is the image of the hamiltonian vector
field associated with $H$.
\hfill{$\blacksquare$}

\medskip\noindent
All the above shows that the presented Lagrangian and Hamiltonian
formalisms work well for all algebroids and not only for Lie
algebroids. The fact that the tensor $\zL_\ze$ is Poisson played
no role in the above considerations and Lie brackets have not been
explicitly used.

\section{Examples}

\noindent {\bf a) Generalized geodesics.} The simplest
hyperregular Lagrangian on $E$ is given by a symmetric
positive-definite metric $g$ in $E$. The metric induces an
isomorphism of the vector bundles $\wt{g}:E\ra E^\*$ which, in
turn, induces an isomorphism of corresponding tensor bundles. With
respect to this isomorphism the metric $g$ corresponds to a
`contravariant metric' $G$. In local coordinates:
$$g=g_{ij}(x)e^i_*\ot e^j_*, \quad G=g^{ij}(x)e_i\ot e_j, \quad
g_{ij}g^{jk}=\zd^k_i, \quad \text{and} \quad
L(x,y)=\frac{1}{2}g_{ij}(x)y^iy^j.$$ The Legendre map is given by
$L_{eg}(x,y)=(x^a,g_{ij}(x)y^j)$ and its inverse is
$L^{-1}_{eg}(x,\zx)=(x^a,g^{ij}(x)\zx_j)$.

\medskip\noindent
It is easy to see that the Hamiltonian dynamics $D$ on $E^\*$ is
represented by the vector field
\be D(x,\zx)=\zr^a_i(x)g^{ij}(x)\zx_j\pa_{x^a}+\left(c^l_{ji}(x)g^{js}(x)-
\frac{1}{2}\zs^a_i(x)\frac{\pa g^{sl}}{\pa
x^a}(x)\right)\zx_s\zx_l\pa_{\zx_i}.\ee This is the Hamiltonian vector
field of $\zL_\ze$ with a Hamiltonian
$$H(x,\zx)=\frac{1}{2}g^{ij}(x)\zx_i\zx_j.$$

\medskip\noindent
The equations $(E^1_L)$ and $(E^2_L)$ coincide and read
\beas\dot{x}^a&=&\zr^a_k(x)y^k,\\
\frac{\xd}{\xd t}\left(g_{ik}y^i\right)&=&\left(c^s_{ik}g_{sj}+
\frac{1}{2}\zs^a_k\frac{\pa g_{ij}}{\pa x^a}\right)y^iy^j.\eeas
The last equation can be rewritten in the form
$$\dot{y}^l+\zG^l_{ij}(x)y^iy^j=0,$$ where
\be\label{gamma}\zG^l_{ij}=\frac{1}{2}g^{kl}\left(\zr^a_j\frac{\pa
g_{ik}}{\pa x^a}+\zr^a_i\frac{\pa g_{jk}}{\pa
x^a}-\zs^a_k\frac{\pa g_{ij}}{\pa
x^a}-c^s_{ik}g_{sj}-c^s_{jk}g_{si}\right).\ee The equations
\be\label{geo}\dot{x}^a=\zr^a_k(x)y^k,\qquad
\dot{y}^l+\zG^l_{ij}(x)y^iy^j=0,\ee with $\zG^l_{ij}$ as in
(\ref{gamma}), are generalized geodesic equations, since for the
case $E=\sT M$ in adapted coordinates we have $c^l_{ij}=0$,
$\zr^a_i=\zs^a_i=\zd^a_i$, so they reduce to the standard geodesic
equation with $\zG^l_{ij}$ being the Christoffel symbols of the
Levi-Civita connection associated with the metric $g$. In the
general case of an algebroid (except for skew-symmetric ones) the
notion of Levi-Civita connection is unclear, since the concept of
the torsion is unclear. The `Christoffel symbols' $\zG^l_{ij}$ can
be understood however as the structure constants of the algebroid
lift of the contravariant metric $G$ pushed forward to $E^\*$ by
the diffeomorphism $\wt{g}$. To be more precise, let us observe
first that the algebroid lift $\xd_\sT^{\ze^+}G$ of $G$ with
respect to the adjoint algebroid structure $\ze^+$ is a symmetric
contravariant tensor on $E$ which reads
$$\xd_\sT^{\ze^+}G=\wt{c}^s_{ij}y^s\pa_{y^i}\ot\pa_{y^j}+
\wt{\zr}^a_i(\pa_{y^i}\ot\pa_{x^a}+\pa_{x^a}\ot\pa_{y^i}),
$$
where
\beas\wt{\zr}^a_i&=&-g^{ij}\zr^a_j,\\
\wt{c}^s_{ij}&=&g^{kj}c^i_{ks}+g^{ki}c^j_{ks}-\frac{\pa
g^{ij}}{\pa x^a}\zs^a_s.\eeas
Its push-forward to $E^\*$ is a
symmetric contravariant tensor which turns out to be
$$\wt{G}=\wt{g}_*(\xd_\sT^{\ze^+}G)=-2\zG^l_{ij}\zx_l\pa_{\zx_i}\ot
\pa_{\zx_j}-\zr^a_i(\pa_{\zx_i}\ot\pa_{x^a}+\pa_{x^a}\ot
\pa{\zx_i}),$$ with $\zG^l_{ij}$ as in (\ref{gamma}). One can
produce, say, a left connection $\nabla$ out of $\wt{G}$ by
putting $\nabla=\frac{1}{2}(\zL_\ze-\wt{g}_*(\xd_\sT^{\ze^+}G))$,
i.e. by
\be\label{LC}\nabla_XY=\frac{1}{2}([X,Y]_\ze-\wt{G}(X,Y)).\ee In
the case of a skew-symmetric algebroid $\ze$, the connection
(\ref{LC}) is the Levi-Civita connection of the metric $g$
(uniquely determined like in the standard case, cf. \cite{GU2})
and the symmetrization of its Christoffel symbols gives exactly
the symbols $\zG^l_{ij}$.

\medskip\noindent {\bf b) Generalized Wong equations.}

\medskip
\noindent Consider an algebroid $E$ which is the direct product
$E=\sT M\ti\g$ of the canonical Lie algebroid $\sT M$ over a
manifold $M$ of dimension $m$ and an arbitrary $\R$-algebra $\g$
of dimension $n$ (the setting could be more general, based on a
short exact sequence of algebroids, but we have chosen this one
for simplicity). Both anchors coincide with the projection $pr_1$
on the first factor, i.e. on $\sT M$. For local coordinates
$(x^a)$ in $M$ and a basis $(v_j)$ of $\g$ we have the adapted
coordinates in $E$: $(x^a,\dx^b,\bv^i)$, where $(\bv^i)$ is the
basis in $\g^\*$ dual to $(v_i)$. Of course, here we understand
$\sT M$ and $M\times\g$ as subbundles of $E$ according to the
natural immersions $I_1$ and $I_2$. Let us assume additionally
that we have a Riemannian metric $g$ on $M$ and a metric $h$ on
$\g$. With every connection $A:\sT M\ra E$, i.e. with every vector
bundle morphism $A:\sT M\ra E$ over the identity such that
$pr_1\circ A=id_{\sT M}$, we can associate a metric $g_A$ on $E$
by
\be\label{metric} g_A(X,Y)=g\left(pr_1X,pr_1Y\right)+h\left(X-A(pr_1X),
Y-A(pr_1Y)\right).\ee
In other words, this is the product of
metrics $g$ and $h$ with respect to the identification of $E$ with
$\sT M\ti\g$ via
$$\bar{A}:\sT M\ti\g\ra\sT M\ti\g,\quad
\bar{A}=A\circ pr_1+I_2\circ pr_2.$$ The metric induces a
hyperregular quadratic Lagrangian $L_A$ on $E$ as above and thus
the corresponding Euler-Lagrange equations associated with the
product algebroid structure $\ze$.

An equivalent approach is to consider, instead of (\ref{metric}),
just the product metric on $\sT M\ti\g$ with the corresponding
Lagrangian of the form \be\label{lag}
L(x,\dot{x},\bar{v})=\frac{1}{2}\left(\sum_{i,j\le
n}h_{ij}\bv^i\bv^j +\sum_{a,b\le m}g_{ab}(x)\dx^a\dx^b\right),\ee
together with a deformed algebroid structure $\ze_A$ on $E=\sT
M\ti\g$ obtained via the isomorphism $\bar{A}:\sT M\ti\g\ra\sT
M\ti\g$. Let us fix this new algebroid structure on $E$ by fixing
a connection $A$. In local coordinates,
$A(\pa_{x^a})=\pa_{x^a}+\sum_{i\le n}A^i_a(x)v_i$. The
contravariant tensor $\zL_{\ze_A}$ on $E^\*\simeq\sT^\* M\ti\g^\*$
reads then
$$\zL_{\ze_A}=C^l_{ij}v_l\pa_{v_i}\ot\pa_{v_j}+A^i_a(x)C^l_{ij}v_l\pa_{p_a}\ot
\pa_{v_j}+A^j_a(x)C^l_{ij}v_l\pa_{v_i}\ot\pa_{p_a}+F^l_{ij}(x)\pa_{p_a}\ot\pa_{p_b}
+\pa_{p_a}\wedge\pa_{x^a},$$ where $C^l_{ij}$ are the structure
constants of the algebra $\g$ with respect to the basis $v_i$, and
$F^l_{ab}$ are the coordinates of the curvature $F$ of the
connection $A$ with respect to the product algebroid structure
$\ze$ defined by $F(X,Y)=[A(X),A(Y)]_\ze-A([X,Y]_\ze)$. In local
coordinates,
$$ F^l_{ab}(x)=\frac{\pa A^l_b}{\pa x^a}(x)-\frac{\pa A^l_a}{\pa
x^b}(x)+A^i_a(x)A^j_b(x)C^l_{ij}.$$ It is a matter of standard
calculations to show that in this case the Hamiltonian dynamics
$D$ on $\sT^\*M\ti\g^\*$ is represented by the equations
\bea\frac{\xd}{\xd t}(x^a)&=&g^{ab}(x)p_b,\nn\\
\frac{\xd}{\xd t}(p_a)&=&F^l_{ba}(x)g^{bc}(x)p_cv_l-\frac{1}{2}\frac{\pa g^{bc}}
{\pa x^a}(x)p_bp_c+A^s_a(x)C^l_{js}h^{ji}v_iv_j,\label{el}\\
\frac{\xd}{\xd t}(v_i)&=&A^j_a(x)C^l_{ji}g^{ab}(x)p_bv_l+C^l_{ji}h^{js}v_lv_s.\nn\eea
Since the Lagrangian (\ref{lag}) is hyperregular, this phase dynamics is Hamiltonian
for the tensor $\zL_{\ze_A}$ and a clear Hamiltonian function reads
$$H=\frac{1}{2}(g^{ab}p_ap_b+h^{ij}v_iv_j).$$ If $\g$ is a Lie
algebra and the metric $h$ is invariant, then
$C^l_{ji}h^{js}v_lv_s=0$ and the above equations reduce to the
celebrated Wong equations \cite{Wo}, usually written in the form
\bea\label{first1}\frac{\xd}{\xd t}(v_i)&=&A^j_a(x)C^l_{ji}\dx^bv_l,\\
\frac{\xd}{\xd t}(p_a)&=&F^l_{ba}(x)\dx^bv_l-\frac{1}{2}\frac{\pa
g^{bc}} {\pa x^a}(x)p_bp_c.\label{second}\eea Possible differences
in signs in various references are effects of differences in
conventions. The equation (\ref{first1}) is usually called the
{\it first Wong equation} and the equation (\ref{second}) -- the
{\it second Wong equation}. Note that these equations have an
alternative Hamiltonian description \cite{AGMM}. The corresponding
`Euler-Lagrange equation' in our general case can be easily
derived from (\ref{el}):
\bea\frac{\xd}{\xd t}(x^a)&=&\dx^a,\nn\\
\frac{\xd}{\xd t}(\dx^d)&=&g^{da}\left(F^l_{ba}h_{ls}\dx^b\bv^s+
A^s_aC^l_{js}h_{ls}\bv^s\bv^j +\frac{1}{2}\left(\frac{\pa
g_{ac}}{\pa x^b}+\frac{\pa g_{bc}}{\pa x^a}-\frac{\pa g_{ab}}{\pa
x^c}\right)\dx^b\dx^c\right),\label{wong}\\
\frac{\xd}{\xd
t}(\bv^k)&=&h^{ik}\left(A^j_aC^l_{ji}h_{ls}\dx^a\bv^s+
C^l_{ji}h_{ls}\bv^j\bv^s\right).\nn\eea The constants $C^l_{ji}$
can be arbitrary in our approach. Of course, the generalized Wong
equations (\ref{wong}) form a particular case of the generalized
geodesic equations (\ref{geo}). Note that Lie algebroid versions
of these equations are known and can be found e.g. in \cite{CM}
and \cite{LMM}.

\bigskip\noindent

\bigskip
\noindent Katarzyna Grabowska\\
Division of Mathematical Methods in Physics \\
                University of Warsaw \\
                Ho\.za 69, 00-681 Warszawa, Poland \\
                 {\tt konieczn@fuw.edu.pl} \\\\
\noindent Janusz Grabowski\\Polish Academy of Sciences\\Institute
of Mathematics\\\'Sniadeckich 8, P.O. Box 21, 00-956 Warszawa,
Poland\\{\tt jagrab@impan.gov.pl}\\\\
\noindent Pawe\l\ Urba\'nski\\
Division of Mathematical Methods in Physics \\
                University of Warsaw \\
                Ho\.za 69, 00-681 Warszawa, Poland \\
                 {\tt urba\'nski@fuw.edu.pl}

\end{document}